\documentclass[journal=nalefd,manuscript=letter,layout=twocolumn]{achemso}

\usepackage{amsmath}
\usepackage[utf8]{inputenc}
\usepackage[T1]{fontenc}
\usepackage{float}
\usepackage{stfloats}
\usepackage{lipsum}
\usepackage[separate-uncertainty=true]{siunitx}
\usepackage[numbers]{natbib}
\usepackage{cuted}
\usepackage{color, soul}

\author{A. Iorio}
\email{andreaiorio@me.com}
\affiliation{Dipartimento di Fisica, Università di Pisa, Largo Bruno Pontecorvo 3, I-56127 Pisa, Italy}
\affiliation{NEST, Istituto Nanoscienze-CNR and Scuola Normale Superiore, I-56127 Pisa, Italy}
\author{M. Rocci}
\affiliation{NEST, Istituto Nanoscienze-CNR and Scuola Normale Superiore, I-56127 Pisa, Italy}
\alsoaffiliation{Current address: Francis Bitter Magnet Laboratory and Plasma Science and Fusion Center, Massachusetts Institute of Technology, Cambridge, MA 02139, USA}
\author{L. Bours}
\affiliation{NEST, Istituto Nanoscienze-CNR and Scuola Normale Superiore, I-56127 Pisa, Italy}
\author{M. Carrega}
\affiliation{NEST, Istituto Nanoscienze-CNR and Scuola Normale Superiore, I-56127 Pisa, Italy}
\author{V. Zannier}
\affiliation{NEST, Istituto Nanoscienze-CNR and Scuola Normale Superiore, I-56127 Pisa, Italy}
\author{L. Sorba}
\affiliation{NEST, Istituto Nanoscienze-CNR and Scuola Normale Superiore, I-56127 Pisa, Italy}
\author{S. Roddaro}
\affiliation{NEST, Istituto Nanoscienze-CNR and Scuola Normale Superiore, I-56127 Pisa, Italy}
\alsoaffiliation{Dipartimento di Fisica, Università di Pisa, Largo Bruno Pontecorvo 3, I-56127 Pisa, Italy}
\author{F. Giazotto}
\affiliation{NEST, Istituto Nanoscienze-CNR and Scuola Normale Superiore, I-56127 Pisa, Italy}
\author{E. Strambini}
\affiliation{NEST, Istituto Nanoscienze-CNR and Scuola Normale Superiore, I-56127 Pisa, Italy}
\email{elia.strambini@sns.it}

\title{Vectorial control of the spin-orbit interaction in suspended InAs nanowires}

\begin{document}
\begin{tocentry}
\includegraphics{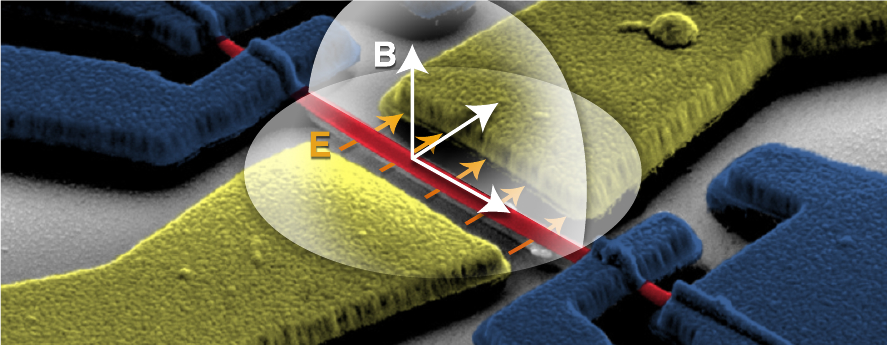}
\end{tocentry}
\begin{strip}
\begin{abstract}
Semiconductor nanowires featuring strong spin-orbit interactions (SOI), represent a promising platform for a broad range of novel technologies, such as spintronic applications or topological quantum computation. However, experimental studies into the nature and the orientation of the SOI vector in these wires remain limited despite being of upmost importance. Typical devices feature the nanowires placed on top of a substrate which modifies the SOI vector and spoils the intrinsic symmetries of the system. In this work, we report experimental results on suspended InAs nanowires, in which the wire symmetries are fully preserved and clearly visible in transport measurements. Using a vectorial magnet, the non-trivial evolution of weak anti-localization (WAL) is tracked through all 3D space, and both the spin-orbit length $l_{SO}$ and coherence length $l_\varphi$ are determined as a function of the magnetic field magnitude and direction. Studying the angular maps of the WAL signal, we demonstrate that the average SOI within the nanowire is isotropic and that our findings are consistent with a semiclassical quasi-1D model of WAL adapted to include the geometrical constraints of the nanostructure. Moreover, by acting on properly designed side gates, we apply an external electric field introducing an additional vectorial Rashba spin-orbit component whose strength can be controlled by external means. These results give important hints on the intrinsic nature of suspended nanowire and can be interesting for the field of spintronics as well as for the manipulation of Majorana bound states in devices based on hybrid semiconductors.
\end{abstract}
\end{strip}

Over the past decades there has been a growing interest in the study of the spin-orbit interactions in semiconductor systems motivated by the possibility of spintronics applications and quantum computing~\cite{golovach_electric-dipole-induced_2006, nowack_coherent_2007, manchon_new_2015, desrat_anticrossings_2005, bercioux_quantum_2015}. Systems with strong SOI offer an ideal platform to develop circuits for the coherent manipulations of electron spins which has led to novel and efficient control schemes that can be applied to make spin transistors or spin-orbit qubits~\cite{datta_electronic_1990, nadj-perge_spinorbit_2010, van_den_berg_fast_2013,bercioux_quantum_2015}. Furthermore, SOI is one of the key ingredients for inducing exotic states of matter, such as topological superconductivity~\cite{kitaev_unpaired_2001, oreg_helical_2010, deng_anomalous_2012, deng_parity_2014,aguado_majorana_2017}. In such systems, a semiconductor nanowire with strong spin-orbit coupling is proximized by a superconductor~\cite{de_gennes_boundary_1964, doh_tunable_2005, krogstrup_epitaxy_2015}, and is predicted to host Majorana zero modes when a Zeeman field is applied orthogonal to the spin-orbit vector~\cite{alicea_new_2012, beenakker_search_2013,tiira_magnetically-driven_2017, virtanen_majorana_2017, bommer_spin-orbit_2018, aguado_majorana_2017}. These modes, exhibiting a non-Abelian exchange statistics, may be used to encode topologically protected qubits~\cite{lutchyn_majorana_2010, oreg_helical_2010,aguado_majorana_2017}.

Despite the vast interest and the variety of new exotic phenomena based on SOI, experimental evidences regarding its origin in semiconductor nanowires and the vectorial dependence of the spin-orbit coupling are limited. Indeed, in quasi-1D systems, the electron properties are strongly affected by the surface states and the confinement potentials that depend on the nanowire geometry and position with respect to the underlying substrate~\cite{bringer_spin_2011, nadj-perge_spectroscopy_2012, wang_anisotropic_2018}. Confinement can strongly enhance the electron gyromagnetic ratio~\cite{roddaro_strong_2007, vaitiekenas_effective_2017}, and the contact between the nanowire and substrate is believed to induce an asymmetry in the confinement potential. In turn, this is the main source of the spin-orbit coupling, acting via the \emph{Rashba effect}~\cite{bychkov_oscillatory_1984, mourik_signatures_2012, bercioux_quantum_2015, wojcik_tuning_2018, campos_spin-orbit_2018, gmitra_first-principles_2016,faria_junior_realistic_2016,kammermeier_spin_2018}. Such an asymmetry causes the pinning of the spin-orbit field in the plane of the substrate, as demonstrated for nanowire quantum dots and for Majorana nanowires~\cite{nadj-perge_spectroscopy_2012, mourik_signatures_2012,montemurro_suspended_2015}. Additionally, the lack of spatial inversion symmetry in the crystal structure can introduce another spin-orbit contribution, namely the \emph{Dresselhaus SOI}~\cite{dresselhaus_spin-orbit_1955,wang_anisotropic_2018,campos_spin-orbit_2018, gmitra_first-principles_2016,faria_junior_realistic_2016,kammermeier_spin_2018}. Furthermore, local internal electric fields, caused by the Fermi level pinning at the nanowire surface, can also give extra Rashba contributions~\cite{bringer_spin_2011,kammermeier_spin_2018}. 

\begin{figure}
\includegraphics{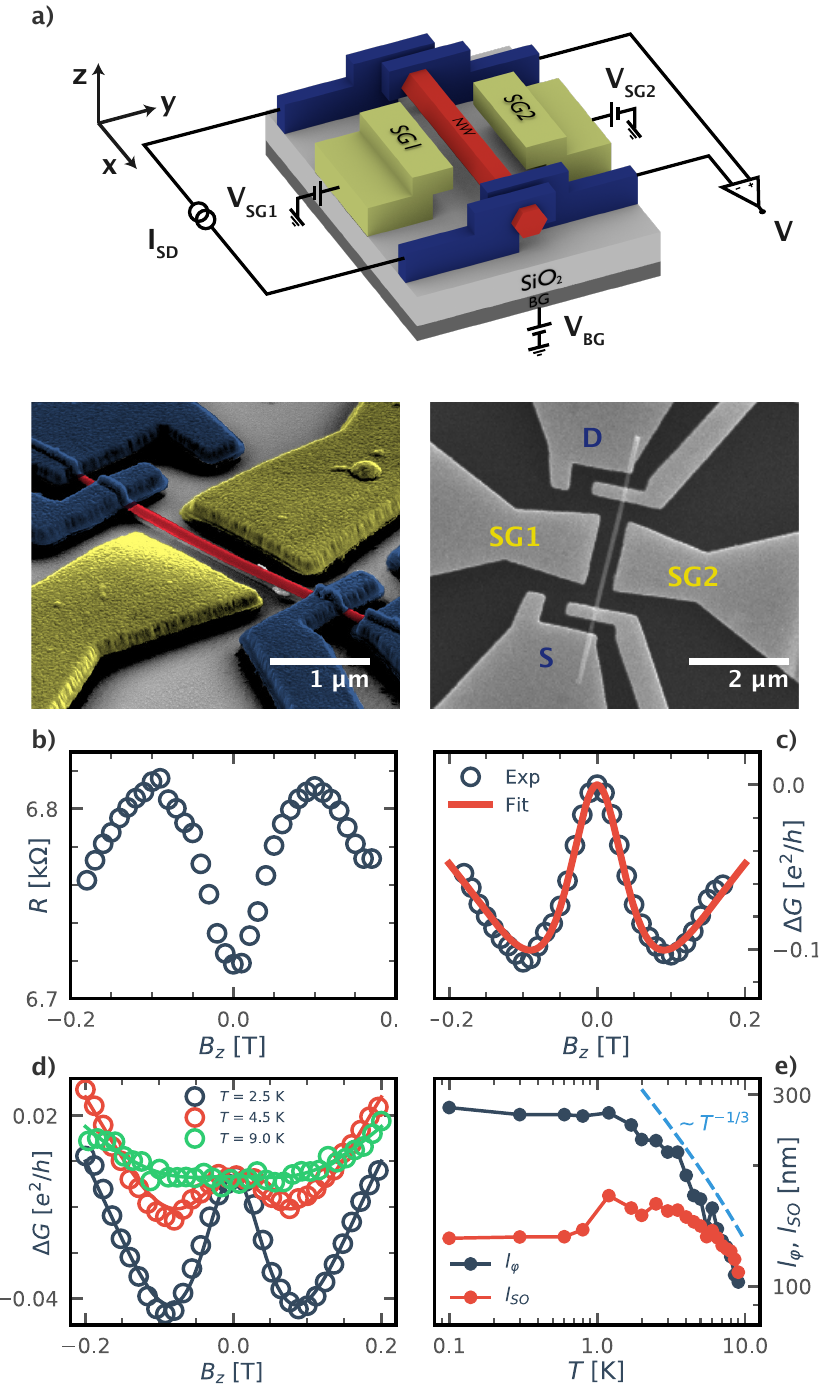}
\caption{a) Top: schematic illustration of a suspended InAs nanowire with a sketch of the four-terminal measurement setup. The nanowire is parallel to the $x$-axis and the substrate is in the $x-y$ plane. Bottom: tilted false-color view and top view SEM images of a representative device. The two side gates $SG1$ and $SG2$ are used to induce electric fields inside the nanowire while keeping its carrier density constant. b, c) Example of a WAL trace taken with the magnetic field along the $z$-axis. In b) the measured resistance $R$ at $T=\SI{50}{\milli\kelvin}$; in c) the correction $\Delta G(B)$ to the conductance extracted from $R$ is fitted with the 1D model expressed in Eq.~(\ref{eq:wal}) giving $l_\varphi \sim \SI{300}{\nm}$ and $l_{SO} \sim \SI{170}{\nm}$. d, e) Temperature dependence of the WAL. In d) WAL curves for three different temperatures are shown. In e) extracted values for $l_{\varphi}$ and $l_{SO}$ as a function of $T$.}
\label{fig:dev}
\end{figure}

In this work, we take a different perspective and investigate the SOI in freely suspended nanowires which are not expected to display any intrinsic electrostatic asymmetry. In this way, we are able to reestablish the natural geometrical degeneracies as the suspended wires offer an ideal platform for studying the intrinsic SOI. The \emph{vectorial} dependence of the SOI on a magnetic field is investigated by tracking the weak anti-localization peak in the magnetoconductance at low temperature (50-200 mK), while rotating the magnetic field. WAL is a dampening of the weak localization (WL), i.e. the localization of charge due to constructive interference of electron waves along time-reversed paths, and it is induced by the presence of spin-orbit coupling~\cite{bergmann_weak_1984, kurdak_quantum_1992, altshuler_magnetoresistance_1981}. While WL leads to a negative correction to the magnetoconductance, the WAL results in a positive one that depends on the electron spin relaxation. From the magnetic field dependence of these two competing effects, the relevant length scales over which spin and phase information are preserved, i.e. spin-orbit length $l_{SO}$ and coherence length $l_\varphi$, can be extracted.

Our typical device is schematically shown in Fig.~\ref{fig:dev}a: a single suspended $n$-type InAs nanowire with two side gate electrodes (SGs), a global back gate (BG) and four ohmic contacts: source (S), drain (D) and two voltage probes. Se-doped InAs nanowires are grown by Au-assisted chemical beam epitaxy on InAs (111) substrates~\cite{gomes_controlling_2015}, resulting in pure defect-free wurtzite nanowires with hexagonal cross section and growth direction corresponding to the $c$-axis. The suspension has been realized via the mechanical transfer of the wire onto a \SI{200}{\nm} thick PMMA layer placed over a SiO$_2$/$n$-Si substrate. The PMMA is then crosslinked at the two nanowire edges while it is dissolved in the middle to provide the suspension. Subsequently, the four ohmic contacts are fabricated by electron beam lithography and Cr/Au evaporation (10/100 nm).

The magnetotransport measurements were performed in a filtered dilution refrigerator at temperatures down to \SI{50}{\milli\kelvin}, using a standard 4-wire lock-in technique. A quasi-DC current excitation of $I_{SD} = \SI{10}{\nA}$ and frequency $f_{SD} = \SI{129.49}{\Hz}$ is used to prevent electron heating effects while a three-axis vectorial magnet is used to investigate the WAL as a function of the magnetic field orientation. Based on transconductance characterizations at $T=\SI{4}{\kelvin}$, we estimate a typical electron concentration $n \simeq \SI{2e18}{\cm^{-3}}$ and mobility $\mu \simeq \SI{1200}{\cm^2/\volt\second}$. The corresponding Fermi velocity $v_F$, mean free path $l_e$ and diffusion coefficient $D = v_F l_e/3$, are evaluated to be $v_F \simeq \SI{2e6}{\m/\s}$, $l_e \simeq \SI{30}{\nm}$ and $D \simeq \SI{200}{\cm^2/s}$ (for more details see the SI).

Due to the finite number of scattering paths in the wire, universal conductance fluctuations (UCF) appear in transport signals and are superimposed to the WAL correction. Since they are of the same magnitude, averaging techniques are needed to recover a clear WAL signal. This is typically achieved by measuring different nanowires connected in parallel~\cite{hansen_spin_2005} or by averaging the conductance $G$ of a single wire on different UCF configurations obtained by gating~\cite{estevez_hernandez_spin-orbit_2010, roulleau_suppression_2010}. In our measurements we employ the latter technique and add a low frequency AC modulation of peak-to-peak amplitude $V^{pp}_{avg} = \SI{6}{\V}$ and frequency $f_{avg} = \SI{10.320}{\Hz}$ to either the side gates or the back gate potential. The signal is integrated over a \SI{1}{\s} time window providing the effective averaged resistance $\langle R(B) \rangle_{V^{pp}_{avg}}$. 

In Fig.~\ref{fig:dev}b a representative, averaged magnetoresistance curve is presented showing the resistance $R(B)$ and the conductance correction $\Delta G(B) = G(B)-G(0)$, with $G(B)=1/R(B)$, measured at $T=\SI{50}{\milli\kelvin}$. The clear peak in the conductance at $B=0$ is due to the presence of a non-vanishing spin relaxation mechanism inside the nanowire. In order to extract the spin and phase relaxation lengths, the correction to the conductivity is studied and compared to the quasi-1D theory for disordered systems~\cite{kurdak_quantum_1992, altshuler_magnetoresistance_1981}
\begin{equation}
\begin{split}
\Delta G(B) \propto -\frac{2e^2}{hL} &\left[\frac{3}{2} \left( \frac{1}{l_\varphi^2} + \frac{4}{3 l_{SO}^2} + \frac{1}{l_B^2}  \right)^{-1/2} \right. \\ 
&\left. - \frac{1}{2} \left(\frac{1}{l_\varphi^2}+ \frac{1}{l_B^2} \right)^{-1/2} \right],
\end{split}
\raisetag{2\normalbaselineskip}
\label{eq:wal}
\end{equation}
where $B$ is the external magnetic field, $L=\SI{2}{\micro\meter}$ the distance between the contacts, $l_\varphi$ the phase coherence length, $l_{SO}$ the spin relaxation length and $l_B$ the magnetic dephasing length, which depends both on $\mathbf B$ and on the confinement geometry, as discussed below. The above quantities are connected to the corresponding timescales by the relation
\begin{equation}
l_{\varphi, SO, B} = \sqrt{D\tau_{\varphi, SO, B}}.
\end{equation}
The model presented in Eq.~(\ref{eq:wal}) is valid in the \emph{weak-field} limit, \emph{i.e.} $l_m \gg W$, with $l_m = \sqrt{\hbar/eB}$ the magnetic length and $W$ the nanowire diameter, and for \emph{diffusive} transport $l_e \ll W$ fulfilled for the nanowire with $W \simeq \SI{90}{\nano\meter}$. It is worth to note that, within this semiclassical model, the spin relaxation length $l_{SO}$ is assumed to be \emph{isotropic}, then not accounting for any angular dependence of the magnetic field. The only quantity that depends on the field orientation is the magnetic dephasing time $\tau_B$, that can be written as~\cite{altshuler_magnetoresistance_1981}
\begin{equation}
\tau_{B} = C\frac{4l_m^4}{DW^2},
\label{eq:dep1}
\end{equation}
where the factor $C$ accounts for the geometry of the nanowire and the orientation of the magnetic field. When the nanowire cross section is a regular polygon, $C$ is isotropic for magnetic fields perpendicular to the wire ($B_\perp$) and it increases for parallel field ($B_\parallel$). In Table~\ref{tbl:C} we report $C$ calculated for different geometries (see the SI for more details).

\begin{center}
\begin{table}
\begin{tabular}{|c|c|c|}
\hline
Nanowire cross section & $B$ orientation & $C$ \\
\hline
Square & $\perp$ & 1.5 \\
Hexagon & $\perp$ & 1.2 \\
Circle & $\perp$ & 1 \\
Circle & $\parallel$ & 2 \\
\hline
\end{tabular}
\caption{Geometrical factors calculated for different nanowire cross sections and orientation of $B$ with respect to the nanowire. See SI for details.}
\label{tbl:C}
\end{table}
\end{center}

Figure~\ref{fig:dev}c shows the fit to a typical WAL measurement, obtained with the magnetic field applied along the $z$-axis. An empirical prefactor $A=3$ has been included in Eq.~($\ref{eq:wal}$) in order to obtain a satisfactory fit to the data as similarly done for WAL in one~\cite{jespersen_crystal_2018} or two dimensional systems~\cite{studenikin_experimental_2003}. This guarantees satisfactory fits for all magnetic field orientations, temperatures, and gate voltages. The magnetoconductance is acquired for different temperatures from \SIrange{0.1}{10}{\kelvin} as shown in Fig.~\ref{fig:dev}d and \ref{fig:dev}e. As expected, a crossover from WAL to WL is observed when $l_\varphi$ becomes smaller than $l_{SO}$ since, when the spin relaxation is slower than the phase relaxation, the spin components are preserved during the electron motion along time-reversed paths and thus the WL effect is recovered. The phase coherence length increases by decreasing temperature and saturates below \SI{1}{\kelvin}. The good agreement with the $\propto T^{-1/3}$ power law indicates that small electron-electron energy transfers in the 1D system constitute the main dephasing mechanism (\emph{Nyquist dephasing})~\cite{altshuler_chapter_1985, wang_phase-coherent_2015}. The saturation of $l_\varphi$ for temperature smaller than $T \sim \SI{1}{\kelvin}$, can be attributed to the external environment noise~\cite{khavin_decoherence_1998}. The temperature dependence of $l_{SO}$ can be ascribed to phonon-induced spin-relaxation processes in the D'yakonov-Perel' mechanism~\cite{ohno_electron_2000,kainz_temperature_2004, hao_strong_2010, wang_electrical_2017}.

\begin{figure}
\includegraphics{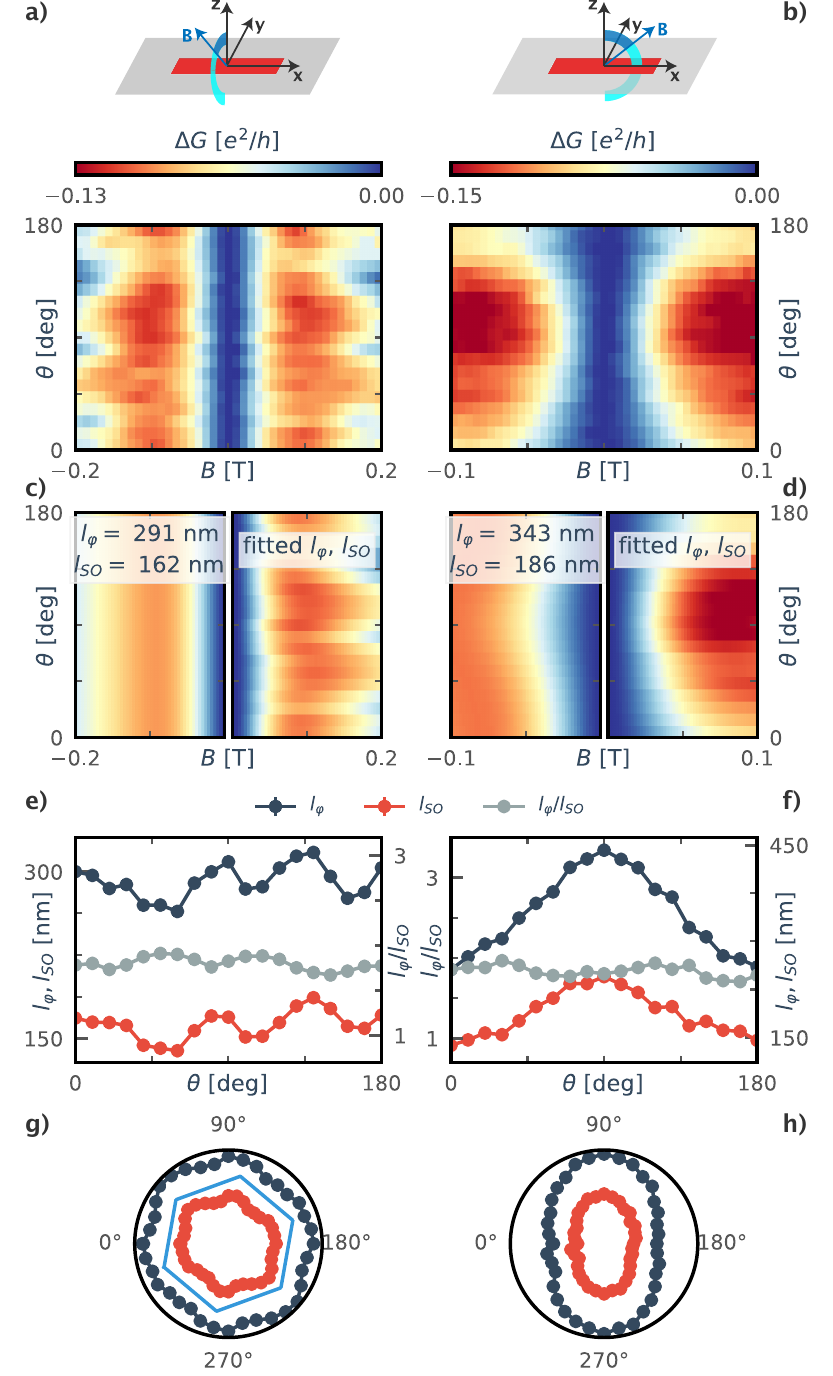}
\caption{Evolution of the WAL in the transverse (left) and longitudinal (right) plane of $\mathbf B$. a, b) A lateral sketch of the nanowire with the measured correction to the conductance $\Delta G$ as a function of the azimuth angle $\theta$ and the magnetic field $B$. c, d) Comparison between theoretical models: on the left, $l_\varphi$ and $l_{SO}$ are fixed. The expression of $\tau_B$ takes into account the different contributions by using $1/\tau_B = \cos^2 \theta/\tau_{B_\perp} + \sin^2 \theta/\tau_{B_\parallel}$~\cite{liang_anisotropic_2010}. On the right, the WAL of a) and b) are fitted with $l_\varphi$ and $l_{SO}$ free to change. e, f) Angle dependence of the fitted relaxation lengths $l_\varphi$ and $l_{SO}$. The strong correlation between the two is made evident by the almost constant ratio $l_\varphi/l_{SO}$. g, h) The same values as above shown in a polar plot. In g) a guide to the eye for the hexagonal geometry rotated by -\SI{10}{\degree} is shown in blue.}
\label{fig:maps}
\end{figure}

The evolution of the WAL peak along the plane orthogonal to the nanowire ($z-y$ plane) is reported in Fig.~\ref{fig:maps}a displaying $\Delta G$ measured vs $B$ and the azimuth angle $\theta$. At low magnetic fields ($|B| < \SI{0.1}{\tesla}$), the WAL peak displays almost no dependence on $\theta$ in agreement with the isotropy of $\tau_B$ expected from the theoretical prediction (as shown in Fig.~\ref{fig:maps}c, left panel). At larger magnetic fields ($|B| \simeq \SI{0.2}{\tesla}$), the WAL shows three clear peaks periodic in $\theta$ (blue regions), suggesting a 6-fold periodicity in \SI{360}{\degree} consistent with the \emph{hexagonal} symmetry of the nanowire cross section. This degeneracy, obtained by the nanowire's suspension, is not easily achieved as small asymmetries in the gates arrangement, or differing charge configurations in different cooling cycles can easily remove it. By fitting the data with Eq.~(\ref{eq:wal}) the same 6-fold periodicity is visible in both $l_{\varphi}$ and $l_{SO}$ as shown in Fig.~\ref{fig:maps}e and \ref{fig:maps}g. At first instance, the modulation of $l_{SO}$ may be ascribed to a vectorial dependence of the spin-orbit coupling induced by the confinement geometry. However, the unexpected correlation between $l_\varphi$ and $l_{SO}$ suggests that this modulation has to be an artifact of the simplified model used for fitting the WAL data. Furthermore, the estimation of $l_\varphi$ should not depend on the orientation of the probing magnetic field. It is important to remark, indeed, that the 1D model of WAL assumes $\tau_{B_\perp}$ to be independent on the field orientation, which seems unlikely for a hexagonal nanowire geometry. The strong correlation and the unphysical behavior of $l_\varphi$ can be resolved by a simple geometrical renormalization of $\tau_B$ in Eq.~(\ref{eq:dep1}), by replacing $W$ with an angle-dependent effective width $W_{eff}(\theta)$. Physically, $W_{eff}$ is the width of the nanowire projected on the plane orthogonal to the magnetic field and accounts for the maximum area available for time-reversed paths. By considering this geometrical correction, with $W_{eff}$ spanning from the diameter to the face-to-face distance of the hexagonal cross section, it is possible to accurately fit the data while obtaining $l_\varphi$ and $l_{SO}$ almost constant to the values $l_\varphi \simeq \SI{310}{\nm}$ and $l_{SO} \simeq \SI{180}{\nm}$ (see SI). Since the hexagonal 6-fold symmetry becomes visible for $|B| > \SI{0.1}{\tesla}$, this correction induced by the geometry is relevant going beyond the weak-field limit (valid for $|B|<\SI{0.1}{\tesla}$). The observed degeneracy in the transverse plane excludes the presence of a favored direction for spin-relaxation. This is in agreement with a Rashba effect that originates from the electric fields due to the nanowire confining potential. The observed degeneracy is also compatible with the symmetry of the Dresselhaus interaction in this plane, but its contribution is expected to be negligible for our nanowire growth direction~\cite{campos_spin-orbit_2018}.

The symmetry of the nanowire in the WAL features is even more evident in the longitudinal planes. Here, a clear 2-fold modulation of the WAL is observed also at low magnetic field as shown in Fig.~\ref{fig:maps}b for the $z-x$ plane. A similar result has been observed as well in the $x-y$ plane (see SI). The evolution of WAL in $\theta$ shows a considerably narrower peak for magnetic fields parallel to the wire axis. Remarkably, this is in contrast with the theoretical predictions, in which a broadening of the WAL peak is expected due to the angular dependence of $\tau_B$ (see left panel of Fig.~\ref{fig:maps}d). By fitting the WAL with Eq.~(\ref{eq:wal}), we observe a periodic modulation of both $l_\varphi$ and $l_{SO}$ covering a range from \SIrange{300}{450}{\nm} for $l_\varphi$ and from \SIrange{150}{200}{\nm} for $l_{SO}$ (Fig.~\ref{fig:maps}f and \ref{fig:maps}h). Again, the strong correlation between the two relaxation lengths and the unexpected trend of $l_\varphi$ suggest a renormalization of $\tau_B$ rather than a real vectorial dependence of the spin-orbit coupling. In this case, the renormalization can be justified by mechanisms leading to a increased/decreased flux pick-up and is opposite to the geometrical scaling of $C$ presented in Table \ref{tbl:C}. Indeed, for perpendicular magnetic fields, \emph{flux-cancellation} phenomena can bring to an enhancement of $\tau_{B_\perp}$~\cite{beenakker_boundary_1988, blomers_electronic_2011}, while, for parallel field, winding trajectories can increase the effective area encircled by time-reversed paths leading to a reduced $\tau_{B_\parallel}$~\cite{scheer_angular_1997}. These mechanisms are particularly relevant when transport is dominated by electron states at the tubular surface of the nanowire, as expected in the case of our suspended nanostructures~\cite{jespersen_probing_2015, jespersen_crystal_2018}. Differently, the substrate typically induces an asymmetric electron distribution in the case of non-suspended nanowires~\cite{degtyarev_features_2017} and indeed previous works showed instead a widening of the WAL shape for magnetic fields parallel to the nanowire~\cite{roulleau_suppression_2010} or no significant magnetic field orientation dependence~\cite{van_weperen_spin-orbit_2015}. 

\begin{figure*}[t]
\centering
\includegraphics{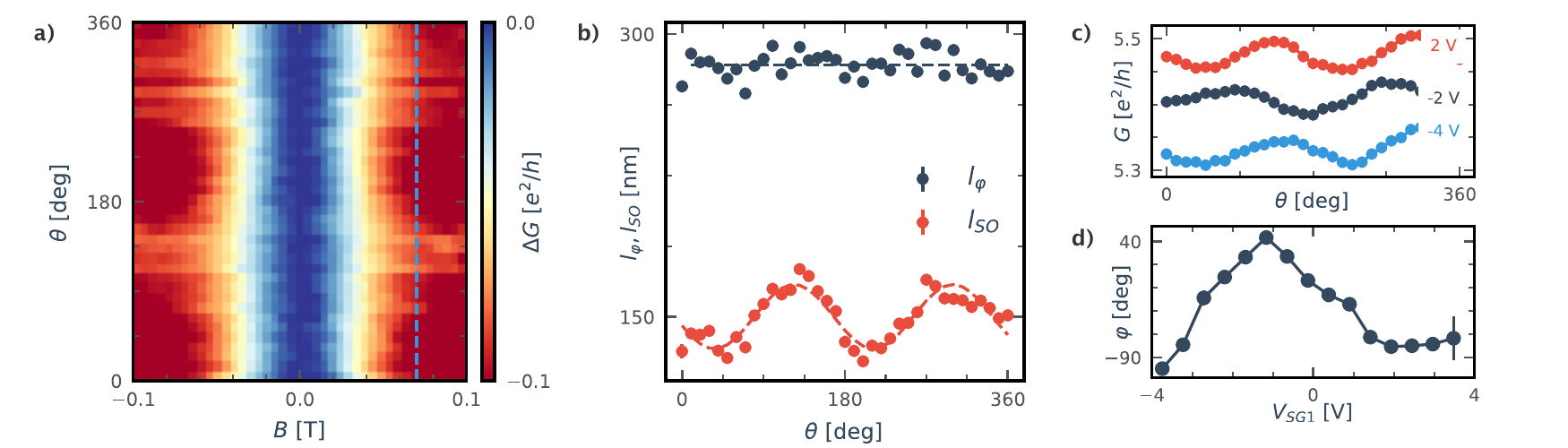}
\caption{a) WAL along the $z-y$ plane in the presence of an external electric field applied along the $\hat y$ direction by setting the asymmetric potential $V_{SG1} = \alpha V_{SG2} = -\SI{4}{\volt}$ to the SGs. b) $l_\varphi$ and $l_{SO}$ obtained by fitting the data in a) as a function of $\theta$. c) Traces of $G(\theta)$ at $B=\SI{70}{\milli\tesla}$ for different $V_{SG}$ values. d) The resulting phase shift $\varphi$ of the traces in c) is shown as a function of the electric field strength inside the nanowire.}
\label{fig:mod}
\end{figure*}

To remove this degeneracy and induce a real vectorial spin-orbit coupling, an external electric field $\mathbf E$ is applied along the $\hat y$ direction. In order to keep the electron density constant inside the wire, the field is induced by applying an asymmetric potential $V_{SG1} = \alpha V_{SG2}$ to the side gate electrodes, with $\alpha = -0.4$ to compensate for an asymmetric capacitive coupling of the two lateral gates and maintain a constant transconductance~\cite{scherubl_electrical_2016}. This technique does not a priori exclude a reshaping of the charge distribution that, as discussed before, could also induce an artificial modulation of both $l_\varphi$ and $l_{SO}$ as frequently observed in similar nanowires~\cite{scherubl_electrical_2016, wang_electrical_2017}. To distinguish between geometrical effects and real modulation of the SOI, we characterize the magnetoconductance in the full $z-y$ plane and in the presence of the external electric field as shown in Fig.~\ref{fig:mod}a. Differently from what reported in Fig.~\ref{fig:maps}a, now the WAL shows a clear 2-fold modulation in $\theta$ and, by fitting the data with Eq.~(\ref{eq:wal}), a variation of $l_{SO}$ is obtained and is totally uncorrelated to the unperturbed $l_\varphi$ (see Fig.~\ref{fig:mod}b). The resulting anisotropy of $l_{SO}$ indicates a vectorial enhancement of the Rashba coupling acting via the external electric field and whose maximal modulation is observed for $\mathbf B$ perpendicular to the direction of $\mathbf E$. To track the evolution of the WAL for different electric field strengths, we directly measured the conductance $G(\theta)$ at $B=\SI{70}{\milli\tesla}$. In Fig.~\ref{fig:mod}c the traces obtained for different side gate voltages are shown. A clear 2-fold periodicity is always observed and a continuous phase shift $\varphi$, shown in Fig.~\ref{fig:mod}d, remarks a vectorial evolution of the SOI. $\varphi$ is related to the angle for which $l_{SO}$ is maximally reduced due to the enhanced Rashba coupling. Remarkably, a saturation towards $-\SI{90}{\degree}$ is observed for side gate voltages reaching $\pm \SI{4}{\volt}$ in agreement with Fig.~\ref{fig:mod}a. For smaller SG voltages the angular evolution is non-trivial and probably related to geometrical effects of the charge distribution or to a residual asymmetry of the confining electric field tilted with respect to the external one.

In conclusion, we have studied the vectorial dependence of the SOI via angle-dependent magnetoconductance in suspended InAs nanowires. The WAL clearly follows the symmetries of the hexagonal nanowire showing a 6-fold degeneracy for $\mathbf B$ lying in the transverse plane and a 2-fold degeneracy in the longitudinal one. These features can be consistently described by the semiclassical quasi-1D model educated to account the geometrical constraints of the system and demonstrate the isotropy of the intrinsic SOI in suspended nanowires. The hexagonal degeneracy is removed in the presence of an external electric field induced by gating. This allows to superimpose a vectorial Rashba coupling inside the nanowire, clearly demonstrated by a 2-fold periodic angular modulation of $l_{SO}$ while $l_\varphi$ remains constant. Our findings show the WAL as practical tool to investigate the vectorial nature and amplitude of the spin-orbit coupling, discriminating between geometrical artifacts and real enhancement of the SOI. The achieved vectorial control of the Rashba SOI is essential for any spintronic application and can inspire novel schemes for the manipulation of Majorana bound states.

\begin{acknowledgement}
We acknowledge P. Virtanen, A. Braggio, G. C. La Rocca, M. Governale and F. Taddei for fruitful discussions. This research was supported in part by the National Science Foundation under Grant No. NSF PHY17-48958. Partial financial support from the European Union's Seventh Framework Programme (FP7/2007-2013)/ERC Grant 615187-COMANCHE and by the Tuscany Region under the FARFAS 2014 project SCIADRO are acknowledged. M.C. and L.S. acknowledge support from the Quant-EraNet project Supertop. S.R. acknowledges support from QUANTRA financed by the Ministry of Foreign Affairs and International Cooperation.
\end{acknowledgement}

\begin{suppinfo}
Electrical characterization, WAL with renormalized $\tau_B$, magnetic dephasing time for different geometries, temperature dependence.
\end{suppinfo}

\bibliography{paper.bib}

\providecommand{\latin}[1]{#1}
\makeatletter
\providecommand{\doi}
  {\begingroup\let\do\@makeother\dospecials
  \catcode`\{=1 \catcode`\}=2 \doi@aux}
\providecommand{\doi@aux}[1]{\endgroup\texttt{#1}}
\makeatother
\providecommand*\mcitethebibliography{\thebibliography}
\csname @ifundefined\endcsname{endmcitethebibliography}
  {\let\endmcitethebibliography\endthebibliography}{}
\begin{mcitethebibliography}{61}
\providecommand*\natexlab[1]{#1}
\providecommand*\mciteSetBstSublistMode[1]{}
\providecommand*\mciteSetBstMaxWidthForm[2]{}
\providecommand*\mciteBstWouldAddEndPuncttrue
  {\def\EndOfBibitem{\unskip.}}
\providecommand*\mciteBstWouldAddEndPunctfalse
  {\let\EndOfBibitem\relax}
\providecommand*\mciteSetBstMidEndSepPunct[3]{}
\providecommand*\mciteSetBstSublistLabelBeginEnd[3]{}
\providecommand*\EndOfBibitem{}
\mciteSetBstSublistMode{f}
\mciteSetBstMaxWidthForm{subitem}{(\alph{mcitesubitemcount})}
\mciteSetBstSublistLabelBeginEnd
  {\mcitemaxwidthsubitemform\space}
  {\relax}
  {\relax}

\bibitem[Golovach \latin{et~al.}(2006)Golovach, Borhani, and
  Loss]{golovach_electric-dipole-induced_2006}
Golovach,~V.~N.; Borhani,~M.; Loss,~D. Electric-dipole-induced spin resonance
  in quantum dots. \emph{Physical Review B} \textbf{2006}, \emph{74},
  165319\relax
\mciteBstWouldAddEndPuncttrue
\mciteSetBstMidEndSepPunct{\mcitedefaultmidpunct}
{\mcitedefaultendpunct}{\mcitedefaultseppunct}\relax
\EndOfBibitem
\bibitem[Nowack \latin{et~al.}(2007)Nowack, Koppens, Nazarov, and
  Vandersypen]{nowack_coherent_2007}
Nowack,~K.~C.; Koppens,~F. H.~L.; Nazarov,~Y.~V.; Vandersypen,~L. M.~K.
  Coherent {Control} of a {Single} {Electron} {Spin} with {Electric} {Fields}.
  \emph{Science} \textbf{2007}, \emph{318}, 1430--1433\relax
\mciteBstWouldAddEndPuncttrue
\mciteSetBstMidEndSepPunct{\mcitedefaultmidpunct}
{\mcitedefaultendpunct}{\mcitedefaultseppunct}\relax
\EndOfBibitem
\bibitem[Manchon \latin{et~al.}(2015)Manchon, Koo, Nitta, Frolov, and
  Duine]{manchon_new_2015}
Manchon,~A.; Koo,~H.~C.; Nitta,~J.; Frolov,~S.~M.; Duine,~R.~A. New
  perspectives for {Rashba} spin{\textendash}orbit coupling. \emph{Nature
  Materials} \textbf{2015}, \emph{14}, 871--882\relax
\mciteBstWouldAddEndPuncttrue
\mciteSetBstMidEndSepPunct{\mcitedefaultmidpunct}
{\mcitedefaultendpunct}{\mcitedefaultseppunct}\relax
\EndOfBibitem
\bibitem[Desrat(2005)]{desrat_anticrossings_2005}
Desrat,~W. Anticrossings of spin-split {Landau} levels in an {InAs}
  two-dimensional electron gas with spin-orbit coupling. \emph{Physical Review
  B} \textbf{2005}, \emph{71}\relax
\mciteBstWouldAddEndPuncttrue
\mciteSetBstMidEndSepPunct{\mcitedefaultmidpunct}
{\mcitedefaultendpunct}{\mcitedefaultseppunct}\relax
\EndOfBibitem
\bibitem[Bercioux and Lucignano()Bercioux, and
  Lucignano]{bercioux_quantum_2015}
Bercioux,~D.; Lucignano,~P. Quantum transport in Rashba spin–orbit materials:
  a review. \emph{Reports on Progress in Physics} \emph{78}, 106001\relax
\mciteBstWouldAddEndPuncttrue
\mciteSetBstMidEndSepPunct{\mcitedefaultmidpunct}
{\mcitedefaultendpunct}{\mcitedefaultseppunct}\relax
\EndOfBibitem
\bibitem[Datta and Das(1990)Datta, and Das]{datta_electronic_1990}
Datta,~S.; Das,~B. Electronic analog of the electro-optic modulator.
  \emph{Applied Physics Letters} \textbf{1990}, \emph{56}, 665--667\relax
\mciteBstWouldAddEndPuncttrue
\mciteSetBstMidEndSepPunct{\mcitedefaultmidpunct}
{\mcitedefaultendpunct}{\mcitedefaultseppunct}\relax
\EndOfBibitem
\bibitem[Nadj-Perge \latin{et~al.}(2010)Nadj-Perge, Frolov, Bakkers, and
  Kouwenhoven]{nadj-perge_spinorbit_2010}
Nadj-Perge,~S.; Frolov,~S.~M.; Bakkers,~E. P. a.~M.; Kouwenhoven,~L.~P.
  Spin{\textendash}orbit qubit in a semiconductor nanowire. \emph{Nature}
  \textbf{2010}, \emph{468}, 1084\relax
\mciteBstWouldAddEndPuncttrue
\mciteSetBstMidEndSepPunct{\mcitedefaultmidpunct}
{\mcitedefaultendpunct}{\mcitedefaultseppunct}\relax
\EndOfBibitem
\bibitem[van~den Berg \latin{et~al.}(2013)van~den Berg, Nadj-Perge, Pribiag,
  Plissard, Bakkers, Frolov, and Kouwenhoven]{van_den_berg_fast_2013}
van~den Berg,~J. W.~G.; Nadj-Perge,~S.; Pribiag,~V.~S.; Plissard,~S.~R.;
  Bakkers,~E. P. A.~M.; Frolov,~S.~M.; Kouwenhoven,~L.~P. Fast {Spin}-{Orbit}
  {Qubit} in an {Indium} {Antimonide} {Nanowire}. \emph{Physical Review
  Letters} \textbf{2013}, \emph{110}, 066806\relax
\mciteBstWouldAddEndPuncttrue
\mciteSetBstMidEndSepPunct{\mcitedefaultmidpunct}
{\mcitedefaultendpunct}{\mcitedefaultseppunct}\relax
\EndOfBibitem
\bibitem[Kitaev(2001)]{kitaev_unpaired_2001}
Kitaev,~A.~Y. Unpaired {Majorana} fermions in quantum wires.
  \emph{Physics-Uspekhi} \textbf{2001}, \emph{44}, 131\relax
\mciteBstWouldAddEndPuncttrue
\mciteSetBstMidEndSepPunct{\mcitedefaultmidpunct}
{\mcitedefaultendpunct}{\mcitedefaultseppunct}\relax
\EndOfBibitem
\bibitem[Oreg \latin{et~al.}(2010)Oreg, Refael, and von
  Oppen]{oreg_helical_2010}
Oreg,~Y.; Refael,~G.; von Oppen,~F. Helical {Liquids} and {Majorana} {Bound}
  {States} in {Quantum} {Wires}. \emph{Physical Review Letters} \textbf{2010},
  \emph{105}, 177002\relax
\mciteBstWouldAddEndPuncttrue
\mciteSetBstMidEndSepPunct{\mcitedefaultmidpunct}
{\mcitedefaultendpunct}{\mcitedefaultseppunct}\relax
\EndOfBibitem
\bibitem[Deng \latin{et~al.}()Deng, Yu, Huang, Larsson, Caroff, and
  Xu]{deng_anomalous_2012}
Deng,~M.~T.; Yu,~C.~L.; Huang,~G.~Y.; Larsson,~M.; Caroff,~P.; Xu,~H.~Q.
  Anomalous Zero-Bias Conductance Peak in a Nb–{InSb} Nanowire–Nb Hybrid
  Device. \emph{Nano Letters} \emph{12}, 6414--6419\relax
\mciteBstWouldAddEndPuncttrue
\mciteSetBstMidEndSepPunct{\mcitedefaultmidpunct}
{\mcitedefaultendpunct}{\mcitedefaultseppunct}\relax
\EndOfBibitem
\bibitem[Deng \latin{et~al.}()Deng, Yu, Huang, Larsson, Caroff, and
  Xu]{deng_parity_2014}
Deng,~M.~T.; Yu,~C.~L.; Huang,~G.~Y.; Larsson,~M.; Caroff,~P.; Xu,~H.~Q. Parity
  independence of the zero-bias conductance peak in a nanowire based
  topological superconductor-quantum dot hybrid device. \emph{Scientific
  Reports} \emph{4}, 7261\relax
\mciteBstWouldAddEndPuncttrue
\mciteSetBstMidEndSepPunct{\mcitedefaultmidpunct}
{\mcitedefaultendpunct}{\mcitedefaultseppunct}\relax
\EndOfBibitem
\bibitem[Aguado()]{aguado_majorana_2017}
Aguado,~R. Majorana quasiparticles in condensed matter. \emph{Riv.Nuovo Cim.}
  \emph{40}, 1\relax
\mciteBstWouldAddEndPuncttrue
\mciteSetBstMidEndSepPunct{\mcitedefaultmidpunct}
{\mcitedefaultendpunct}{\mcitedefaultseppunct}\relax
\EndOfBibitem
\bibitem[DE~GENNES(1964)]{de_gennes_boundary_1964}
DE~GENNES,~P.~G. Boundary {Effects} in {Superconductors}. \emph{Reviews of
  Modern Physics} \textbf{1964}, \emph{36}, 225--237\relax
\mciteBstWouldAddEndPuncttrue
\mciteSetBstMidEndSepPunct{\mcitedefaultmidpunct}
{\mcitedefaultendpunct}{\mcitedefaultseppunct}\relax
\EndOfBibitem
\bibitem[Doh \latin{et~al.}(2005)Doh, Dam, Roest, Bakkers, Kouwenhoven, and
  Franceschi]{doh_tunable_2005}
Doh,~Y.-J.; Dam,~J. A.~v.; Roest,~A.~L.; Bakkers,~E. P. A.~M.;
  Kouwenhoven,~L.~P.; Franceschi,~S.~D. Tunable {Supercurrent} {Through}
  {Semiconductor} {Nanowires}. \emph{Science} \textbf{2005}, \emph{309},
  272--275\relax
\mciteBstWouldAddEndPuncttrue
\mciteSetBstMidEndSepPunct{\mcitedefaultmidpunct}
{\mcitedefaultendpunct}{\mcitedefaultseppunct}\relax
\EndOfBibitem
\bibitem[Krogstrup \latin{et~al.}(2015)Krogstrup, Ziino, Chang, Albrecht,
  Madsen, Johnson, Nyg{\r a}rd, Marcus, and Jespersen]{krogstrup_epitaxy_2015}
Krogstrup,~P.; Ziino,~N. L.~B.; Chang,~W.; Albrecht,~S.~M.; Madsen,~M.~H.;
  Johnson,~E.; Nyg{\r a}rd,~J.; Marcus,~C.~M.; Jespersen,~T.~S. Epitaxy of
  semiconductor{\textendash}superconductor nanowires. \emph{Nature Materials}
  \textbf{2015}, \emph{14}, 400--406\relax
\mciteBstWouldAddEndPuncttrue
\mciteSetBstMidEndSepPunct{\mcitedefaultmidpunct}
{\mcitedefaultendpunct}{\mcitedefaultseppunct}\relax
\EndOfBibitem
\bibitem[Alicea(2012)]{alicea_new_2012}
Alicea,~J. New directions in the pursuit of {Majorana} fermions in solid state
  systems. \emph{Reports on Progress in Physics} \textbf{2012}, \emph{75},
  076501\relax
\mciteBstWouldAddEndPuncttrue
\mciteSetBstMidEndSepPunct{\mcitedefaultmidpunct}
{\mcitedefaultendpunct}{\mcitedefaultseppunct}\relax
\EndOfBibitem
\bibitem[Beenakker(2013)]{beenakker_search_2013}
Beenakker,~C. Search for {Majorana} {Fermions} in {Superconductors}.
  \emph{Annual Review of Condensed Matter Physics} \textbf{2013}, \emph{4},
  113--136\relax
\mciteBstWouldAddEndPuncttrue
\mciteSetBstMidEndSepPunct{\mcitedefaultmidpunct}
{\mcitedefaultendpunct}{\mcitedefaultseppunct}\relax
\EndOfBibitem
\bibitem[Tiira \latin{et~al.}(2017)Tiira, Strambini, Amado, Roddaro, San-Jose,
  Aguado, Bergeret, Ercolani, Sorba, and
  Giazotto]{tiira_magnetically-driven_2017}
Tiira,~J.; Strambini,~E.; Amado,~M.; Roddaro,~S.; San-Jose,~P.; Aguado,~R.;
  Bergeret,~F.~S.; Ercolani,~D.; Sorba,~L.; Giazotto,~F. Magnetically-driven
  colossal supercurrent enhancement in {InAs} nanowire {Josephson} junctions.
  \emph{Nature Communications} \textbf{2017}, \emph{8}, 14984\relax
\mciteBstWouldAddEndPuncttrue
\mciteSetBstMidEndSepPunct{\mcitedefaultmidpunct}
{\mcitedefaultendpunct}{\mcitedefaultseppunct}\relax
\EndOfBibitem
\bibitem[Virtanen \latin{et~al.}(2018)Virtanen, Bergeret, Strambini, Giazotto,
  and Braggio]{virtanen_majorana_2017}
Virtanen,~P.; Bergeret,~F.~S.; Strambini,~E.; Giazotto,~F.; Braggio,~A.
  Majorana bound states in hybrid two-dimensional Josephson junctions with
  ferromagnetic insulators. \emph{Phys. Rev. B} \textbf{2018}, \emph{98},
  020501\relax
\mciteBstWouldAddEndPuncttrue
\mciteSetBstMidEndSepPunct{\mcitedefaultmidpunct}
{\mcitedefaultendpunct}{\mcitedefaultseppunct}\relax
\EndOfBibitem
\bibitem[Bommer \latin{et~al.}(2018)Bommer, Zhang, G{\"u}l, Nijholt, Wimmer,
  Rybakov, Garaud, Rodic, Babaev, Troyer, Car, Plissard, Bakkers, Watanabe,
  Taniguchi, and Kouwenhoven]{bommer_spin-orbit_2018}
Bommer,~J. D.~S. \latin{et~al.}  Spin-{Orbit} {Protection} of {Induced}
  {Superconductivity} in {Majorana} {Nanowires}. \emph{arXiv:1807.01940
  [cond-mat]} \textbf{2018}, arXiv: 1807.01940\relax
\mciteBstWouldAddEndPuncttrue
\mciteSetBstMidEndSepPunct{\mcitedefaultmidpunct}
{\mcitedefaultendpunct}{\mcitedefaultseppunct}\relax
\EndOfBibitem
\bibitem[Lutchyn \latin{et~al.}(2010)Lutchyn, Sau, and
  Das~Sarma]{lutchyn_majorana_2010}
Lutchyn,~R.~M.; Sau,~J.~D.; Das~Sarma,~S. Majorana {Fermions} and a
  {Topological} {Phase} {Transition} in {Semiconductor}-{Superconductor}
  {Heterostructures}. \emph{Physical Review Letters} \textbf{2010}, \emph{105},
  077001\relax
\mciteBstWouldAddEndPuncttrue
\mciteSetBstMidEndSepPunct{\mcitedefaultmidpunct}
{\mcitedefaultendpunct}{\mcitedefaultseppunct}\relax
\EndOfBibitem
\bibitem[Bringer(2011)]{bringer_spin_2011}
Bringer,~A. Spin precession and modulation in ballistic cylindrical nanowires
  due to the {Rashba} effect. \emph{Physical Review B} \textbf{2011},
  \emph{83}\relax
\mciteBstWouldAddEndPuncttrue
\mciteSetBstMidEndSepPunct{\mcitedefaultmidpunct}
{\mcitedefaultendpunct}{\mcitedefaultseppunct}\relax
\EndOfBibitem
\bibitem[Nadj-Perge \latin{et~al.}(2012)Nadj-Perge, Pribiag, van~den Berg, Zuo,
  Plissard, Bakkers, Frolov, and Kouwenhoven]{nadj-perge_spectroscopy_2012}
Nadj-Perge,~S.; Pribiag,~V.~S.; van~den Berg,~J. W.~G.; Zuo,~K.;
  Plissard,~S.~R.; Bakkers,~E. P. A.~M.; Frolov,~S.~M.; Kouwenhoven,~L.~P.
  Spectroscopy of {Spin}-{Orbit} {Quantum} {Bits} in {Indium} {Antimonide}
  {Nanowires}. \emph{Physical Review Letters} \textbf{2012}, \emph{108},
  166801\relax
\mciteBstWouldAddEndPuncttrue
\mciteSetBstMidEndSepPunct{\mcitedefaultmidpunct}
{\mcitedefaultendpunct}{\mcitedefaultseppunct}\relax
\EndOfBibitem
\bibitem[Wang \latin{et~al.}()Wang, Huang, Huang, Xue, Pan, Zhao, and
  Xu]{wang_anisotropic_2018}
Wang,~J.-Y.; Huang,~G.-Y.; Huang,~S.; Xue,~J.; Pan,~D.; Zhao,~J.; Xu,~H.
  Anisotropic Pauli Spin-Blockade Effect and Spin–Orbit Interaction Field in
  an {InAs} Nanowire Double Quantum Dot. \emph{Nano Letters} \emph{18},
  4741--4747\relax
\mciteBstWouldAddEndPuncttrue
\mciteSetBstMidEndSepPunct{\mcitedefaultmidpunct}
{\mcitedefaultendpunct}{\mcitedefaultseppunct}\relax
\EndOfBibitem
\bibitem[Roddaro \latin{et~al.}(2007)Roddaro, Fuhrer, Fasth, Samuelson, Xiang,
  and Lieber]{roddaro_strong_2007}
Roddaro,~S.; Fuhrer,~A.; Fasth,~C.; Samuelson,~L.; Xiang,~J.; Lieber,~C.~M.
  Strong g-{Factor} {Anisotropy} in {Hole} {Quantum} {Dots} {Defined} in
  {Ge}/{Si} {Nanowires}. \emph{arXiv:0706.2883 [cond-mat]} \textbf{2007},
  arXiv: 0706.2883\relax
\mciteBstWouldAddEndPuncttrue
\mciteSetBstMidEndSepPunct{\mcitedefaultmidpunct}
{\mcitedefaultendpunct}{\mcitedefaultseppunct}\relax
\EndOfBibitem
\bibitem[Vaitiek{\.e}nas \latin{et~al.}(2017)Vaitiek{\.e}nas, Deng, Nyg{\r
  a}rd, Krogstrup, and Marcus]{vaitiekenas_effective_2017}
Vaitiek{\.e}nas,~S.; Deng,~M.~T.; Nyg{\r a}rd,~J.; Krogstrup,~P.; Marcus,~C.~M.
  Effective g-factor in {Majorana} {Wires}. \emph{arXiv:1710.04300 [cond-mat]}
  \textbf{2017}, arXiv: 1710.04300\relax
\mciteBstWouldAddEndPuncttrue
\mciteSetBstMidEndSepPunct{\mcitedefaultmidpunct}
{\mcitedefaultendpunct}{\mcitedefaultseppunct}\relax
\EndOfBibitem
\bibitem[Bychkov and Rashba(1984)Bychkov, and Rashba]{bychkov_oscillatory_1984}
Bychkov,~Y.~A.; Rashba,~E.~I. Oscillatory effects and the magnetic
  susceptibility of carriers in inversion layers. \emph{Journal of Physics C:
  Solid State Physics} \textbf{1984}, \emph{17}, 6039\relax
\mciteBstWouldAddEndPuncttrue
\mciteSetBstMidEndSepPunct{\mcitedefaultmidpunct}
{\mcitedefaultendpunct}{\mcitedefaultseppunct}\relax
\EndOfBibitem
\bibitem[Mourik \latin{et~al.}(2012)Mourik, Zuo, Frolov, Plissard, Bakkers, and
  Kouwenhoven]{mourik_signatures_2012}
Mourik,~V.; Zuo,~K.; Frolov,~S.~M.; Plissard,~S.~R.; Bakkers,~E. P. a.~M.;
  Kouwenhoven,~L.~P. Signatures of {Majorana} {Fermions} in {Hybrid}
  {Superconductor}-{Semiconductor} {Nanowire} {Devices}. \emph{Science}
  \textbf{2012}, \emph{336}, 1003--1007\relax
\mciteBstWouldAddEndPuncttrue
\mciteSetBstMidEndSepPunct{\mcitedefaultmidpunct}
{\mcitedefaultendpunct}{\mcitedefaultseppunct}\relax
\EndOfBibitem
\bibitem[Wójcik \latin{et~al.}()Wójcik, Bertoni, and
  Goldoni]{wojcik_tuning_2018}
Wójcik,~P.; Bertoni,~A.; Goldoni,~G. Tuning Rashba spin-orbit coupling in
  homogeneous semiconductor nanowires. \emph{Physical Review B} \emph{97},
  165401\relax
\mciteBstWouldAddEndPuncttrue
\mciteSetBstMidEndSepPunct{\mcitedefaultmidpunct}
{\mcitedefaultendpunct}{\mcitedefaultseppunct}\relax
\EndOfBibitem
\bibitem[Campos \latin{et~al.}()Campos, Faria~Junior, Gmitra, Sipahi, and
  Fabian]{campos_spin-orbit_2018}
Campos,~T.; Faria~Junior,~P.~E.; Gmitra,~M.; Sipahi,~G.~M.; Fabian,~J.
  Spin-orbit coupling effects in zinc-blende InSb and wurtzite InAs nanowires:
  Realistic calculations with multiband
  $\mathbf{k}\ifmmode\cdot\else\textperiodcentered\fi{}\mathbf{p}$ method.
  \emph{Physical Review B} \emph{97}, 245402\relax
\mciteBstWouldAddEndPuncttrue
\mciteSetBstMidEndSepPunct{\mcitedefaultmidpunct}
{\mcitedefaultendpunct}{\mcitedefaultseppunct}\relax
\EndOfBibitem
\bibitem[Gmitra and Fabian()Gmitra, and Fabian]{gmitra_first-principles_2016}
Gmitra,~M.; Fabian,~J. First-principles studies of orbital and spin-orbit
  properties of {GaAs}, {GaSb}, {InAs}, and {InSb} zinc-blende and wurtzite
  semiconductors. \emph{Physical Review B} \emph{94}, 165202\relax
\mciteBstWouldAddEndPuncttrue
\mciteSetBstMidEndSepPunct{\mcitedefaultmidpunct}
{\mcitedefaultendpunct}{\mcitedefaultseppunct}\relax
\EndOfBibitem
\bibitem[Faria~Junior \latin{et~al.}()Faria~Junior, Campos, Bastos, Gmitra,
  Fabian, and Sipahi]{faria_junior_realistic_2016}
Faria~Junior,~P.~E.; Campos,~T.; Bastos,~C. M.~O.; Gmitra,~M.; Fabian,~J.;
  Sipahi,~G.~M. Realistic multiband
  $k\ifmmode\cdot\else\textperiodcentered\fi{}p$ approach from ab initio and
  spin-orbit coupling effects of InAs and InP in wurtzite phase. \emph{Physical
  Review B} \emph{93}, 235204\relax
\mciteBstWouldAddEndPuncttrue
\mciteSetBstMidEndSepPunct{\mcitedefaultmidpunct}
{\mcitedefaultendpunct}{\mcitedefaultseppunct}\relax
\EndOfBibitem
\bibitem[Kammermeier \latin{et~al.}()Kammermeier, Wenk, Dirnberger, Bougeard,
  and Schliemann]{kammermeier_spin_2018}
Kammermeier,~M.; Wenk,~P.; Dirnberger,~F.; Bougeard,~D.; Schliemann,~J. Spin
  relaxation in wurtzite nanowires. \emph{Physical Review B} \emph{98},
  035407\relax
\mciteBstWouldAddEndPuncttrue
\mciteSetBstMidEndSepPunct{\mcitedefaultmidpunct}
{\mcitedefaultendpunct}{\mcitedefaultseppunct}\relax
\EndOfBibitem
\bibitem[Montemurro \latin{et~al.}(2015)Montemurro, Stornaiuolo, Massarotti,
  Ercolani, Sorba, Beltram, Tafuri, and Roddaro]{montemurro_suspended_2015}
Montemurro,~D.; Stornaiuolo,~D.; Massarotti,~D.; Ercolani,~D.; Sorba,~L.;
  Beltram,~F.; Tafuri,~F.; Roddaro,~S. Suspended {InAs} nanowire {Josephson}
  junctions assembled via dielectrophoresis. \emph{Nanotechnology}
  \textbf{2015}, \emph{26}, 385302\relax
\mciteBstWouldAddEndPuncttrue
\mciteSetBstMidEndSepPunct{\mcitedefaultmidpunct}
{\mcitedefaultendpunct}{\mcitedefaultseppunct}\relax
\EndOfBibitem
\bibitem[Dresselhaus(1955)]{dresselhaus_spin-orbit_1955}
Dresselhaus,~G. Spin-{Orbit} {Coupling} {Effects} in {Zinc} {Blende}
  {Structures}. \emph{Physical Review} \textbf{1955}, \emph{100},
  580--586\relax
\mciteBstWouldAddEndPuncttrue
\mciteSetBstMidEndSepPunct{\mcitedefaultmidpunct}
{\mcitedefaultendpunct}{\mcitedefaultseppunct}\relax
\EndOfBibitem
\bibitem[Bergmann(1984)]{bergmann_weak_1984}
Bergmann,~G. Weak localization in thin films: a time-of-flight experiment with
  conduction electrons. \emph{Physics Reports} \textbf{1984}, \emph{107},
  1--58\relax
\mciteBstWouldAddEndPuncttrue
\mciteSetBstMidEndSepPunct{\mcitedefaultmidpunct}
{\mcitedefaultendpunct}{\mcitedefaultseppunct}\relax
\EndOfBibitem
\bibitem[Kurdak \latin{et~al.}(1992)Kurdak, Chang, Chin, and
  Chang]{kurdak_quantum_1992}
Kurdak,~{\c C}.; Chang,~A.~M.; Chin,~A.; Chang,~T.~Y. Quantum interference
  effects and spin-orbit interaction in quasi-one-dimensional wires and rings.
  \emph{Physical Review B} \textbf{1992}, \emph{46}, 6846--6856\relax
\mciteBstWouldAddEndPuncttrue
\mciteSetBstMidEndSepPunct{\mcitedefaultmidpunct}
{\mcitedefaultendpunct}{\mcitedefaultseppunct}\relax
\EndOfBibitem
\bibitem[Al'Tshuler and Aronov(1981)Al'Tshuler, and
  Aronov]{altshuler_magnetoresistance_1981}
Al'Tshuler,~B.~L.; Aronov,~A.~G. Magnetoresistance of thin films and of wires
  in a longitudinal magnetic field. \emph{Soviet Journal of Experimental and
  Theoretical Physics Letters} \textbf{1981}, \emph{33}, 499\relax
\mciteBstWouldAddEndPuncttrue
\mciteSetBstMidEndSepPunct{\mcitedefaultmidpunct}
{\mcitedefaultendpunct}{\mcitedefaultseppunct}\relax
\EndOfBibitem
\bibitem[Gomes \latin{et~al.}(2015)Gomes, Ercolani, Zannier, Beltram, and
  Sorba]{gomes_controlling_2015}
Gomes,~U.~P.; Ercolani,~D.; Zannier,~V.; Beltram,~F.; Sorba,~L. Controlling the
  diameter distribution and density of {InAs} nanowires grown by {Au}-assisted
  methods. \emph{Semiconductor Science and Technology} \textbf{2015},
  \emph{30}, 115012\relax
\mciteBstWouldAddEndPuncttrue
\mciteSetBstMidEndSepPunct{\mcitedefaultmidpunct}
{\mcitedefaultendpunct}{\mcitedefaultseppunct}\relax
\EndOfBibitem
\bibitem[Hansen(2005)]{hansen_spin_2005}
Hansen,~A.~E. Spin relaxation in {InAs} nanowires studied by tunable weak
  antilocalization. \emph{Physical Review B} \textbf{2005}, \emph{71}\relax
\mciteBstWouldAddEndPuncttrue
\mciteSetBstMidEndSepPunct{\mcitedefaultmidpunct}
{\mcitedefaultendpunct}{\mcitedefaultseppunct}\relax
\EndOfBibitem
\bibitem[Est{\'e}vez~Hern{\'a}ndez
  \latin{et~al.}(2010)Est{\'e}vez~Hern{\'a}ndez, Akabori, Sladek, Volk, Alagha,
  Hardtdegen, Pala, Demarina, Gr{\"u}tzmacher, and
  Sch{\"a}pers]{estevez_hernandez_spin-orbit_2010}
Est{\'e}vez~Hern{\'a}ndez,~S.; Akabori,~M.; Sladek,~K.; Volk,~C.; Alagha,~S.;
  Hardtdegen,~H.; Pala,~M.~G.; Demarina,~N.; Gr{\"u}tzmacher,~D.;
  Sch{\"a}pers,~T. Spin-orbit coupling and phase coherence in {InAs} nanowires.
  \emph{Physical Review B} \textbf{2010}, \emph{82}, 235303\relax
\mciteBstWouldAddEndPuncttrue
\mciteSetBstMidEndSepPunct{\mcitedefaultmidpunct}
{\mcitedefaultendpunct}{\mcitedefaultseppunct}\relax
\EndOfBibitem
\bibitem[Roulleau \latin{et~al.}(2010)Roulleau, Choi, Riedi, Heinzel,
  Shorubalko, Ihn, and Ensslin]{roulleau_suppression_2010}
Roulleau,~P.; Choi,~T.; Riedi,~S.; Heinzel,~T.; Shorubalko,~I.; Ihn,~T.;
  Ensslin,~K. Suppression of weak antilocalization in {InAs} nanowires.
  \emph{Physical Review B} \textbf{2010}, \emph{81}, 155449\relax
\mciteBstWouldAddEndPuncttrue
\mciteSetBstMidEndSepPunct{\mcitedefaultmidpunct}
{\mcitedefaultendpunct}{\mcitedefaultseppunct}\relax
\EndOfBibitem
\bibitem[Jespersen \latin{et~al.}(2018)Jespersen, Krogstrup, Lunde, Tanta,
  Kanne, Johnson, and Nyg{\r a}rd]{jespersen_crystal_2018}
Jespersen,~T.~S.; Krogstrup,~P.; Lunde,~A.~M.; Tanta,~R.; Kanne,~T.;
  Johnson,~E.; Nyg{\r a}rd,~J. Crystal orientation dependence of the spin-orbit
  coupling in {InAs} nanowires. \emph{Physical Review B} \textbf{2018},
  \emph{97}, 041303\relax
\mciteBstWouldAddEndPuncttrue
\mciteSetBstMidEndSepPunct{\mcitedefaultmidpunct}
{\mcitedefaultendpunct}{\mcitedefaultseppunct}\relax
\EndOfBibitem
\bibitem[Studenikin(2003)]{studenikin_experimental_2003}
Studenikin,~S.~A. Experimental study of weak antilocalization effects in a
  high-mobility In$_x$Ga$_{1-x}$As/InP quantum well. \emph{Physical Review B}
  \textbf{2003}, \emph{68}\relax
\mciteBstWouldAddEndPuncttrue
\mciteSetBstMidEndSepPunct{\mcitedefaultmidpunct}
{\mcitedefaultendpunct}{\mcitedefaultseppunct}\relax
\EndOfBibitem
\bibitem[Altshuler and Aronov(1985)Altshuler, and
  Aronov]{altshuler_chapter_1985}
Altshuler,~B.~L.; Aronov,~A.~G. In \emph{Modern {Problems} in {Condensed}
  {Matter} {Sciences}}; Efros,~A.~L., Pollak,~M., Eds.;
  Electron{\textendash}{Electron} {Interactions} in {Disordered} {Systems};
  Elsevier, 1985; Vol.~10; pp 1--153\relax
\mciteBstWouldAddEndPuncttrue
\mciteSetBstMidEndSepPunct{\mcitedefaultmidpunct}
{\mcitedefaultendpunct}{\mcitedefaultseppunct}\relax
\EndOfBibitem
\bibitem[Wang \latin{et~al.}()Wang, Guo, Kang, Pan, Li, Fan, Zhao, and
  Xu]{wang_phase-coherent_2015}
Wang,~L.~B.; Guo,~J.~K.; Kang,~N.; Pan,~D.; Li,~S.; Fan,~D.; Zhao,~J.;
  Xu,~H.~Q. Phase-coherent transport and spin relaxation in {InAs} nanowires
  grown by molecule beam epitaxy. \emph{Applied Physics Letters} \emph{106},
  173105\relax
\mciteBstWouldAddEndPuncttrue
\mciteSetBstMidEndSepPunct{\mcitedefaultmidpunct}
{\mcitedefaultendpunct}{\mcitedefaultseppunct}\relax
\EndOfBibitem
\bibitem[Khavin(1998)]{khavin_decoherence_1998}
Khavin,~Y.~B. Decoherence and the {Thouless} {Crossover} in {One}-{Dimensional}
  {Conductors}. \emph{Physical Review Letters} \textbf{1998}, \emph{81},
  1066--1069\relax
\mciteBstWouldAddEndPuncttrue
\mciteSetBstMidEndSepPunct{\mcitedefaultmidpunct}
{\mcitedefaultendpunct}{\mcitedefaultseppunct}\relax
\EndOfBibitem
\bibitem[Ohno \latin{et~al.}()Ohno, Terauchi, Adachi, Matsukura, and
  Ohno]{ohno_electron_2000}
Ohno,~Y.; Terauchi,~R.; Adachi,~T.; Matsukura,~F.; Ohno,~H. Electron spin
  relaxation beyond D'yakonov–Perel’ interaction in {GaAs}/{AlGaAs} quantum
  wells. \emph{Physica E: Low-dimensional Systems and Nanostructures} \emph{6},
  817--820\relax
\mciteBstWouldAddEndPuncttrue
\mciteSetBstMidEndSepPunct{\mcitedefaultmidpunct}
{\mcitedefaultendpunct}{\mcitedefaultseppunct}\relax
\EndOfBibitem
\bibitem[Kainz \latin{et~al.}()Kainz, Rössler, and
  Winkler]{kainz_temperature_2004}
Kainz,~J.; Rössler,~U.; Winkler,~R. Temperature dependence of Dyakonov-Perel
  spin relaxation in zinc-blende semiconductor quantum structures.
  \emph{Physical Review B} \emph{70}, 195322\relax
\mciteBstWouldAddEndPuncttrue
\mciteSetBstMidEndSepPunct{\mcitedefaultmidpunct}
{\mcitedefaultendpunct}{\mcitedefaultseppunct}\relax
\EndOfBibitem
\bibitem[Hao \latin{et~al.}()Hao, Tu, Cao, Zhou, Li, Guo, Fung, Ji, Guo, and
  Lu]{hao_strong_2010}
Hao,~X.-J.; Tu,~T.; Cao,~G.; Zhou,~C.; Li,~H.-O.; Guo,~G.-C.; Fung,~W.~Y.;
  Ji,~Z.; Guo,~G.-P.; Lu,~W. Strong and Tunable Spin-Orbit Coupling of
  One-Dimensional Holes in Ge/Si Core/Shell Nanowires. \emph{Nano Letters}
  \emph{10}, 2956--2960\relax
\mciteBstWouldAddEndPuncttrue
\mciteSetBstMidEndSepPunct{\mcitedefaultmidpunct}
{\mcitedefaultendpunct}{\mcitedefaultseppunct}\relax
\EndOfBibitem
\bibitem[Wang \latin{et~al.}(2017)Wang, Deacon, Yao, Lieber, and
  Ishibashi]{wang_electrical_2017}
Wang,~R.; Deacon,~R.~S.; Yao,~J.; Lieber,~C.~M.; Ishibashi,~K. Electrical
  modulation of weak-antilocalization and spin{\textendash}orbit interaction in
  dual gated {Ge}/{Si} core/shell nanowires. \emph{Semiconductor Science and
  Technology} \textbf{2017}, \emph{32}, 094002\relax
\mciteBstWouldAddEndPuncttrue
\mciteSetBstMidEndSepPunct{\mcitedefaultmidpunct}
{\mcitedefaultendpunct}{\mcitedefaultseppunct}\relax
\EndOfBibitem
\bibitem[Liang \latin{et~al.}(2010)Liang, Du, and Gao]{liang_anisotropic_2010}
Liang,~D.; Du,~J.; Gao,~X. P.~A. Anisotropic magnetoconductance of a {InAs}
  nanowire: {Angle}-dependent suppression of one-dimensional weak localization.
  \emph{Physical Review B} \textbf{2010}, \emph{81}, 153304\relax
\mciteBstWouldAddEndPuncttrue
\mciteSetBstMidEndSepPunct{\mcitedefaultmidpunct}
{\mcitedefaultendpunct}{\mcitedefaultseppunct}\relax
\EndOfBibitem
\bibitem[Beenakker and van Houten(1988)Beenakker, and van
  Houten]{beenakker_boundary_1988}
Beenakker,~C. W.~J.; van Houten,~H. Boundary scattering and weak localization
  of electrons in a magnetic field. \emph{Physical Review B} \textbf{1988},
  \emph{38}, 3232--3240\relax
\mciteBstWouldAddEndPuncttrue
\mciteSetBstMidEndSepPunct{\mcitedefaultmidpunct}
{\mcitedefaultendpunct}{\mcitedefaultseppunct}\relax
\EndOfBibitem
\bibitem[Bl{\"o}mers \latin{et~al.}(2011)Bl{\"o}mers, Lepsa, Luysberg,
  Gr{\"u}tzmacher, L{\"u}th, and Sch{\"a}pers]{blomers_electronic_2011}
Bl{\"o}mers,~C.; Lepsa,~M.~I.; Luysberg,~M.; Gr{\"u}tzmacher,~D.; L{\"u}th,~H.;
  Sch{\"a}pers,~T. Electronic {Phase} {Coherence} in {InAs} {Nanowires}.
  \emph{Nano Letters} \textbf{2011}, \emph{11}, 3550--3556\relax
\mciteBstWouldAddEndPuncttrue
\mciteSetBstMidEndSepPunct{\mcitedefaultmidpunct}
{\mcitedefaultendpunct}{\mcitedefaultseppunct}\relax
\EndOfBibitem
\bibitem[Scheer \latin{et~al.}(1997)Scheer, L{\"o}hneysen, Mirlin, W{\"o}lfle,
  and Hein]{scheer_angular_1997}
Scheer,~E.; L{\"o}hneysen,~H.~v.; Mirlin,~A.~D.; W{\"o}lfle,~P.; Hein,~H.
  Angular {Dependence} of {Universal} {Conductance} {Fluctuations} in
  {Noble}-{Metal} {Nanowires}. \emph{Physical Review Letters} \textbf{1997},
  \emph{78}, 3362--3365\relax
\mciteBstWouldAddEndPuncttrue
\mciteSetBstMidEndSepPunct{\mcitedefaultmidpunct}
{\mcitedefaultendpunct}{\mcitedefaultseppunct}\relax
\EndOfBibitem
\bibitem[Jespersen(2015)]{jespersen_probing_2015}
Jespersen,~T.~S. Probing the spatial electron distribution in {InAs} nanowires
  by anisotropic magnetoconductance fluctuations. \emph{Physical Review B}
  \textbf{2015}, \emph{91}\relax
\mciteBstWouldAddEndPuncttrue
\mciteSetBstMidEndSepPunct{\mcitedefaultmidpunct}
{\mcitedefaultendpunct}{\mcitedefaultseppunct}\relax
\EndOfBibitem
\bibitem[Degtyarev \latin{et~al.}(2017)Degtyarev, Khazanova, and
  Demarina]{degtyarev_features_2017}
Degtyarev,~V.~E.; Khazanova,~S.~V.; Demarina,~N.~V. Features of electron gas in
  {InAs} nanowires imposed by interplay between nanowire geometry, doping and
  surface states. \emph{Scientific Reports} \textbf{2017}, \emph{7}, 3411\relax
\mciteBstWouldAddEndPuncttrue
\mciteSetBstMidEndSepPunct{\mcitedefaultmidpunct}
{\mcitedefaultendpunct}{\mcitedefaultseppunct}\relax
\EndOfBibitem
\bibitem[van Weperen \latin{et~al.}(2015)van Weperen, Tarasinski, Eeltink,
  Pribiag, Plissard, Bakkers, Kouwenhoven, and
  Wimmer]{van_weperen_spin-orbit_2015}
van Weperen,~I.; Tarasinski,~B.; Eeltink,~D.; Pribiag,~V.~S.; Plissard,~S.~R.;
  Bakkers,~E. P. A.~M.; Kouwenhoven,~L.~P.; Wimmer,~M. Spin-orbit interaction
  in {InSb} nanowires. \emph{Physical Review B} \textbf{2015}, \emph{91},
  201413\relax
\mciteBstWouldAddEndPuncttrue
\mciteSetBstMidEndSepPunct{\mcitedefaultmidpunct}
{\mcitedefaultendpunct}{\mcitedefaultseppunct}\relax
\EndOfBibitem
\bibitem[Scher{\"u}bl \latin{et~al.}(2016)Scher{\"u}bl, F{\"u}l{\"o}p, Madsen,
  Nyg{\r a}rd, and Csonka]{scherubl_electrical_2016}
Scher{\"u}bl,~Z.; F{\"u}l{\"o}p,~G.; Madsen,~M.~H.; Nyg{\r a}rd,~J.; Csonka,~S.
  Electrical tuning of {Rashba} spin-orbit interaction in multigated {InAs}
  nanowires. \emph{Physical Review B} \textbf{2016}, \emph{94}, 035444\relax
\mciteBstWouldAddEndPuncttrue
\mciteSetBstMidEndSepPunct{\mcitedefaultmidpunct}
{\mcitedefaultendpunct}{\mcitedefaultseppunct}\relax
\EndOfBibitem
\end{mcitethebibliography}

\clearpage
\onecolumn
\section{Supporting Information}

\subsection{Electrical characterization}
\begin{figure}[H]
\includegraphics{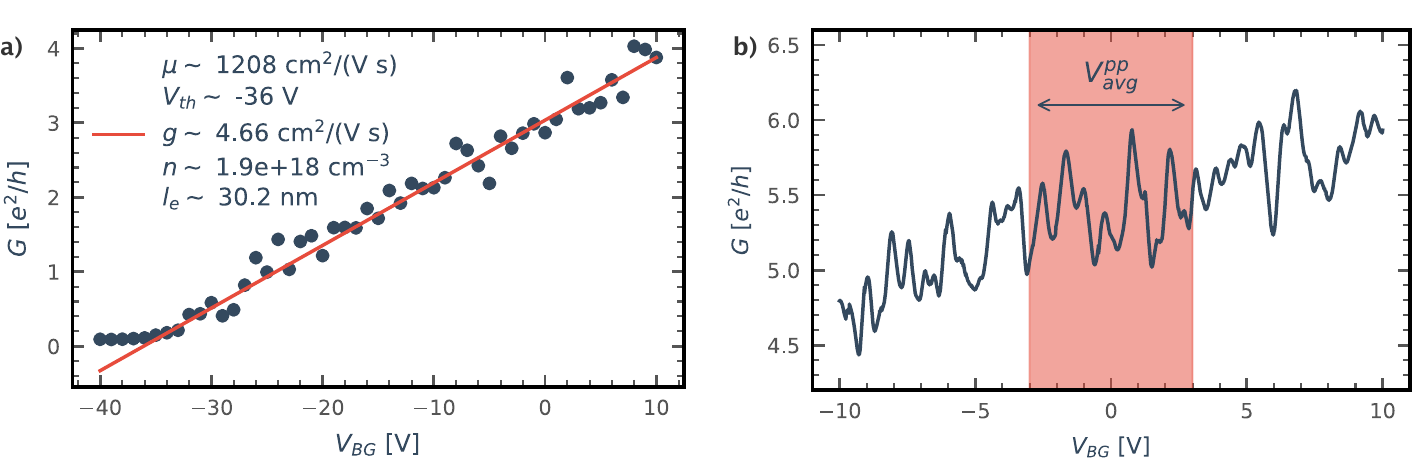}
\caption{a) Transconductance of a typical nanowire measured in the 4-terminal configuration showing the conductance $G$ as a function of the back gate potential $V_{BG}$ at $T=\SI{4}{\kelvin}$. b) Conductance $G$ as a function of the back gate voltage $V_{BG}$ is now shown at $T=\SI{50}{\milli\kelvin}$. In red, the voltage window $V^{pp}_{avg}$ used for averaging the UCF is indicated.}
\label{fig1}
\end{figure}

In order to electrically characterize our nanowires we have performed standard $I-V$ measurements in a four-terminal configuration. We have thus extracted carrier mobility values by fitting the conductance curves $G$ while changing the back gate voltage $V_{BG}$. In Fig.~\ref{fig1}a a typical transconductance obtained at $T=\SI{4}{\kelvin}$ in the four-terminal configuration is shown. The conductance $G$ is fitted with the expression:
\begin{equation}
G(V_{BG}) = \frac{\mu C_{BG}}{L^2}(V_{BG} - V_{th}),
\label{eq:1}
\end{equation}
where $C_{BG} \sim \SI{120}{\atto\farad}$ is the back gate capacitance, as estimated by electrostatic simulation, $L=\SI{2}{\micro\meter}$ is the separation between the contacts, $\mu$ is the nanowire mobility and $V_{th}$ is the threshold voltage. By fitting the data with the expression (\ref{eq:1}) a mobility $\mu \simeq \SI{1200}{\square\cm/\volt\second}$ is estimated from which a typical carrier density at $V_{BG}=\SI{0}{\volt}$ of $n \simeq \SI{2e+18}{\per\cubic\cm}$ is obtained. From the electron relaxation time $\tau_e = \frac{m^* \mu}{e}$ and the Fermi velocity $v_F = \frac{\hbar}{m^*}(3\pi^2 n)^{1/3}$, we can then evaluate a mean free path $l_e = v_F \tau_e \simeq \SI{30}{\nm}$, where the effective mass $m^* = 0.023 m_e$ for InAs has been used.

In Fig.~\ref{fig1}b a the conductance $G$ as a function of the back gate voltage $V_{BG}$ is shown at a lower temperature of $T=\SI{50}{\milli\kelvin}$. At this temperature, UCF are clearly visible and superimposed to the conductance signal. In order to average their contribution in the magnetoconductance measurements, an AC back gate modulation is used as explained in the main text. The voltage window of $V^{pp}_{avg} = \SI{6}{\volt}$ used for averaging is indicated in red.

\begin{figure}[H]
\includegraphics{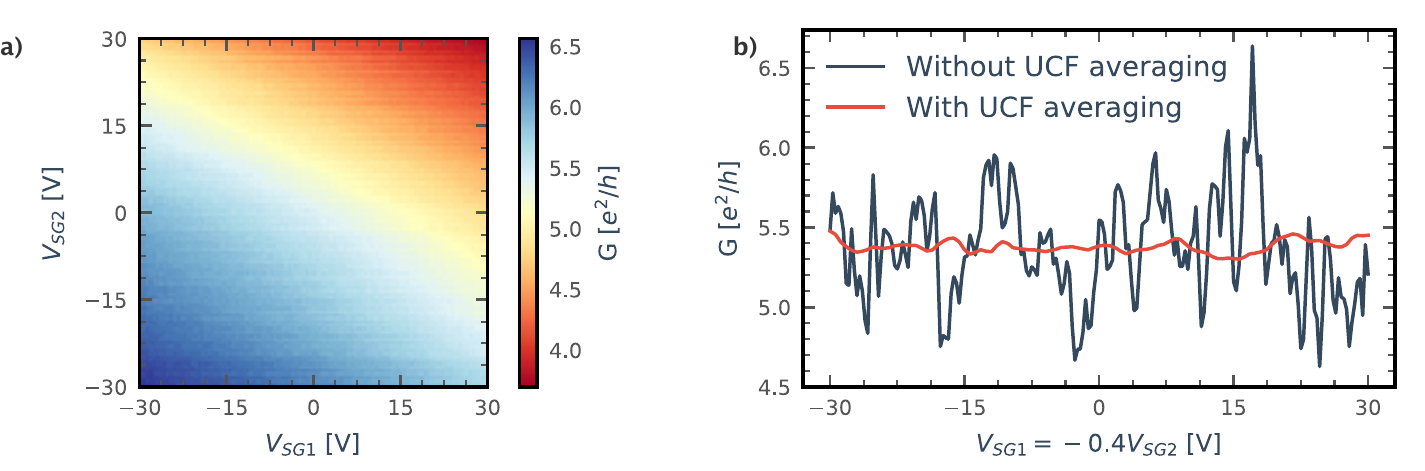}
\caption{a) The conductance $G$ is shown as a function of the two side gates voltage $V_{SG1}$ and $V_{SG2}$ in a color plot. b) The conductance $G$ as a function of the two SG voltages asymmetrically swept. In blue the curve obtained without UCF averaging is shown, while in red the AC back gate potential is turned on, averaging the conductance fluctuations.}
\label{fig2}
\end{figure}

To fully characterize our device, we have also estimated the capacitive coupling of both side gates with respect to the nanowire as shown in Fig.~\ref{fig2}a. This allowed us to precisely control its carrier density and to induce tunable electric fields without affecting the overall wire charge concentration. In Fig.~\ref{fig2}b the conductance $G$ is shown while sweeping the side gate voltages by applying the asymmetric potential $V_{SG1}=-0.4V_{SG2}$. In this way we achieved to keep constant the nanowire conductance $G$ while applying the external electric field. The curve obtained when no averaging is performed is indicated in blue, in which UCF are still clearly visible, while in red the curve obtained when the back gate modulation is turned on in which the UCF contribution is completely averaged. 

\subsection{Weak anti-localization with renormalized $\tau_B$}
\begin{figure}[H]
\includegraphics{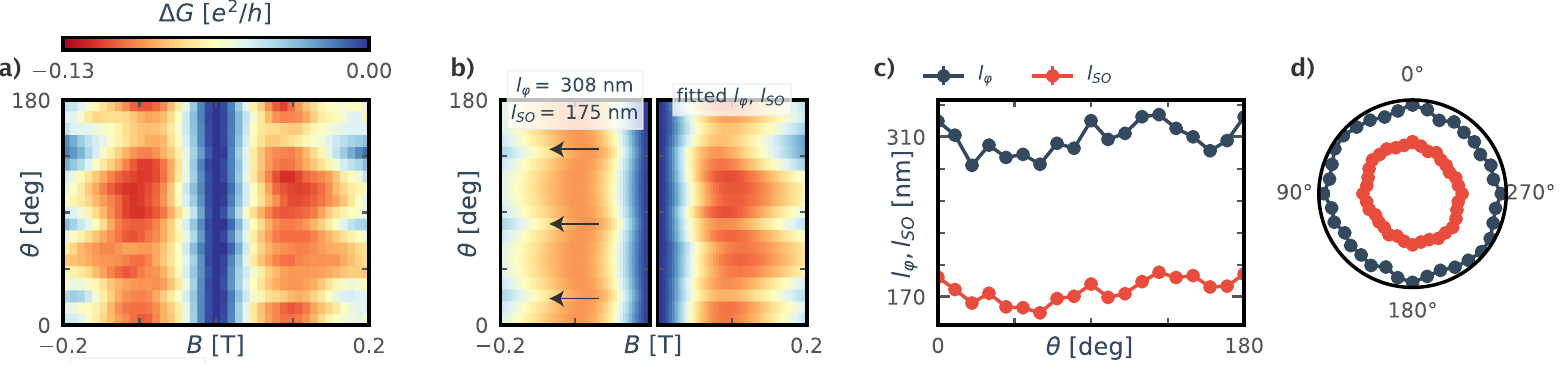}
\caption{The WAL measurements in the $z-y$ plane shown a) are now fitted with the modified expression of $\tau_B(\theta)$ in b). The resulting $l_\varphi$ and $l_{SO}$ shown in c) and d) are now almost flat while changing the magnetic field orientation.}
\label{fig3}
\end{figure}

Figure \ref{fig3}a shows the WAL investigated in the $z-y$ plane as a function of the magnetic field orientation as discussed in the main text. In order to fit the data, here the magnetic dephasing time $\tau_B$ is modified while changing the field orientation by spanning $W$ from the diameter $W_C$ at $\theta=0$ (i.e. the corner-to-corner distance of the hexagonal cross section) to the face-to-face distance at $\theta = \pi/2$:
\begin{equation}
\tau_B = C \frac{4l_M^2}{DW(\theta)^2} \qquad W(\theta) = \frac{W_C\sqrt{3}}{2}\frac{1}{\cos[\pi/6 - \theta~(\text{mod}~\pi/3)]}.
\end{equation}
We note that within the semiclassical quasi-1D model of WAL~\cite{beenakker_boundary_1988, altshuler_magnetoresistance_1981}, $\tau_B$ is predicted to be independent of the magnetic field orientation in the longitudinal plane. The fit of the data with the modified expression of $\tau_B(\theta)$ is shown in Fig.~\ref{fig3}b. The 6-fold periodicity visible at high magnetic field is now well described by the modified expression of $\tau_B$ and is reflected in the fitted values for $l_\varphi$ and $l_{SO}$ that remain constant around the values $l_\varphi \sim \SI{310}{\nano\meter}$ and $l_{SO} \sim \SI{180}{\nano\meter}$.

\begin{figure}[H]
\includegraphics{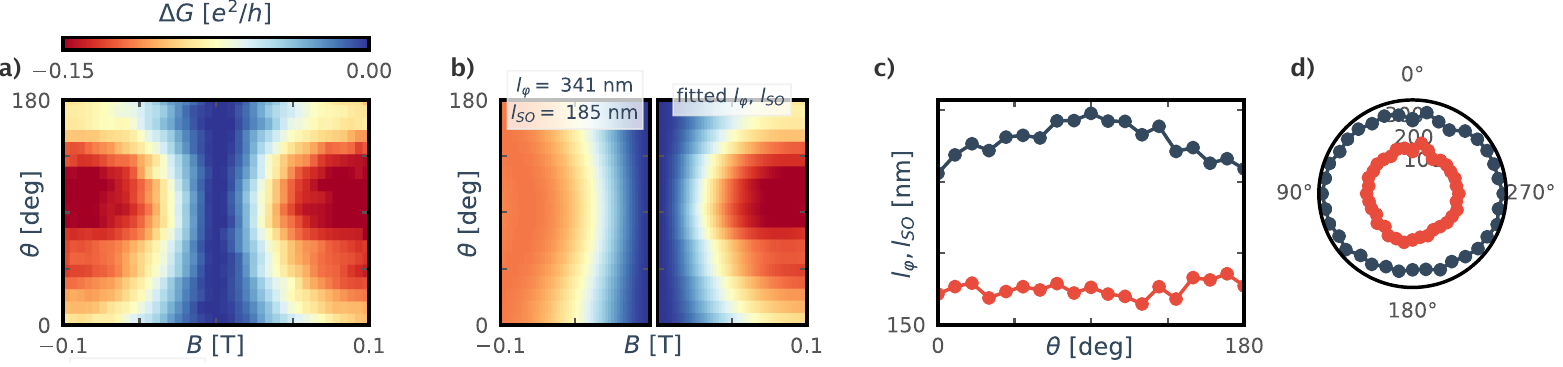}
\caption{The WAL in the $z-x$ plane in a) is now fitted in b) by switching the values of $\tau_{B_\perp}$ and $\tau_{B_\parallel}$. In c) and d) the resulting $l_\varphi$ and $l_{SO}$ are now constant and uncorrelated without showing any unphysical angular modulation.}
\label{fig4}
\end{figure}

As discussed in the main text, the WAL measurements in the $z-x$ plane (Fig.~\ref{fig4}a) show an opposite behavior with respect to the simple geometrical expectations. However, as already argued, different mechanisms of flux pick-up can bring to a reduction/enhancement of the values of $\tau_{B_{\perp}}$ and $\tau_{B_{\parallel}}$. We have thus adapted the semiclassical model to properly take into account these mechanisms by renormalizing their values: in Fig.~\ref{fig4}b, the data are now fitted by simply switching the values of the two magnetic dephasing times.

\begin{figure}[H]
\includegraphics{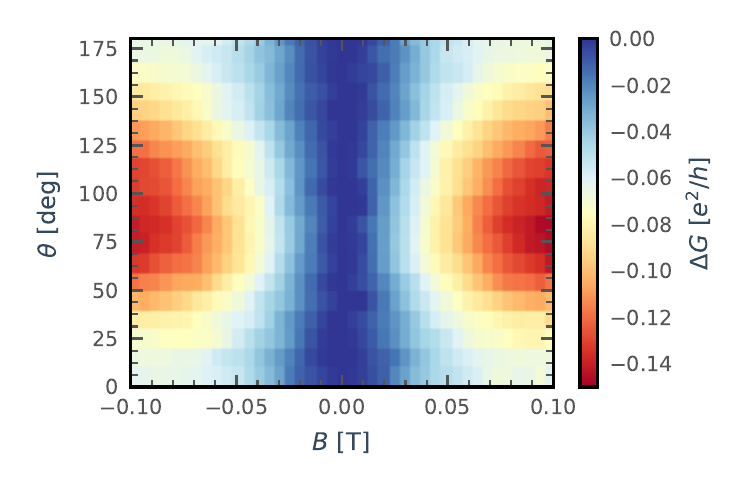}
\caption{WAL measured with magnetic field along the $x-y$ plane.}
\label{fig5}
\end{figure}

Similarly to the transverse $z-x$ plane, also in the $x-y$ plane a widening of the WAL peak is observed for magnetic field parallel to the nanowire (Fig.~\ref{fig5}).

\subsection{Magnetic dephasing time for different geometries}
\begin{figure}[H]
\includegraphics{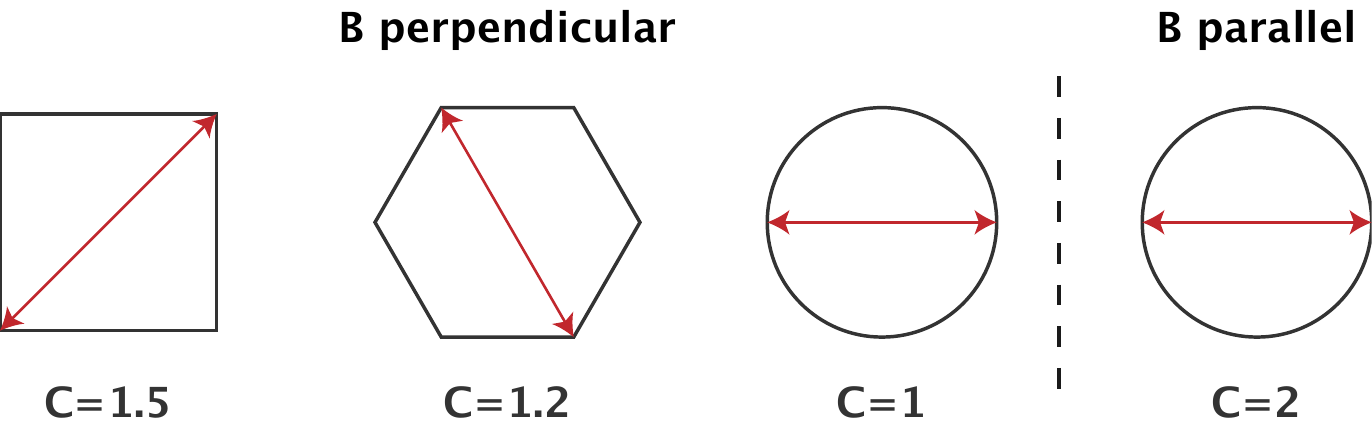}
\caption{Different nanowire cross sections with the corresponding diameter $W$. The value of $C$ obtained by the semiclassical theory is indicated.}
\label{fig6}
\end{figure}

Within the semiclassical quasi-1D model, the expression of the magnetic dephasing time $\tau_B$ in the diffusive regime is essentially related to the mean value of the vector potential $\mathbf A$ over the volume $V$ of the sample~\cite{altshuler_magnetoresistance_1981,beenakker_boundary_1988}:

\begin{equation}
\tau_B = \frac{\hbar^2}{4e^2}\frac{1}{D}\frac{1}{\langle\mathbf A^2\rangle_V},
\end{equation}
if a proper gauge is chosen, such as $\mathbf A \cdot \hat n = 0$ where $\hat n$ is the normal to the surface of the sample. For magnetic field $\mathbf B$ perpendicular to the wire axis, we can choose $\mathbf A = (0,0,f(x,y))$, e.g.:
\begin{align*}
\text{if } \mathbf B = B\hat x &\quad\Rightarrow\quad \mathbf A = (0,0,By), \\	
\text{if } \mathbf B = B\hat y &\quad\Rightarrow\quad \mathbf A = (0,0,-Bx), \\	
\text{if } \mathbf B = B(cos\alpha \hat x + \sin\alpha \hat y) &\quad\Rightarrow\quad \mathbf A = (0,0,B\cos\alpha y - B \sin\alpha x).
\end{align*}
We can thus evaluate the average $\langle\mathbf A^2\rangle_V$ over the volume, which yields
\begin{align*}
\langle A^2_\perp \rangle_{V} & = \langle B^2\cos^2y^2 + B^2\sin^2x^2-B^2\sin\alpha\cos\alpha xy \rangle_V =\\
& = B^2 (\cos^2\alpha \langle y^2\rangle_V + \sin^2\alpha \langle x^2\rangle_V -B^2 \sin\alpha\cos\alpha \langle xy \rangle_V) = \\
& = B^2(\cos^2\alpha \langle y^2 \rangle_V + \sin^2\alpha \langle x^2 \rangle_V) = \\
& = B^2 \langle y^2 \rangle_V \qquad 
\text{if } \langle y^2\rangle_V = \langle x^2 \rangle_V,
\end{align*}
where $\langle xy \rangle_V = 0$ and the last equality holds true for geometries with at least one rotational symmetry $\neq 2\pi$.
For magnetic field $\mathbf B$ parallel to the wire axis, with wire of circular cross section, we can use the symmetric gauge $\mathbf A = (-By/2, Bx/2, 0)$ and obtain
\begin{equation}
\langle A^2_\parallel \rangle_V = \frac{B^2}{4} \langle y^2+x^2\rangle_V = \frac{B^2}{2}\langle y^2 \rangle_V = \frac{1}{2} \langle A^2_\perp \rangle_V.
\end{equation}
A different non-trivial gauge should be chosen in order to satisfy $\mathbf A \cdot \hat n = 0$ for non-circular nanowire cross section. By calculating $\langle y^2\rangle_V$ on the desired geometry we can express $\tau_B$ as
\begin{equation}
\tau_B = C\frac{4l_m^4}{DW^2},
\end{equation}
with the values of $C$ reported in Fig.~\ref{fig6}. However, we remark that evaluating the average $\langle y^2\rangle_V$ is not equivalent to replace $W^2$ in the expression of $\tau_B$ with a different cross sectional area as previously done in different works~\cite{liang_anisotropic_2010, scherubl_electrical_2016}.

\subsection{Temperature dependence}
\begin{figure}[H]
\includegraphics{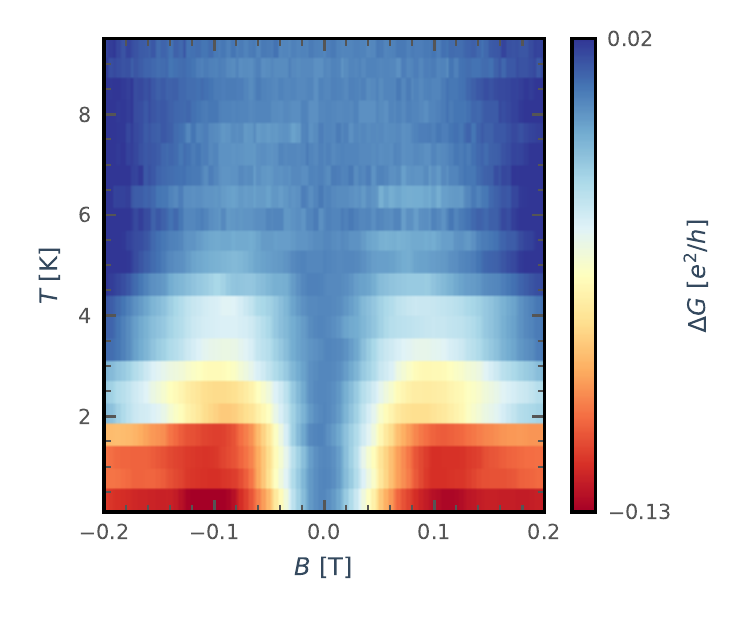}
\caption{Temperature dependence of the WAL signal as a function of the temperature $T$ ranging from \SIrange{0.1}{10}{\K}.}
\label{fig7}
\end{figure}

The magnetoconductance acquired for different temperatures $T$ from \SIrange{0.1}{10}{\K} is shown in Fig.~\ref{fig7}. The clear decrease of the height of the WAL peak as the temperature raises is the due to reduced coherence length. The resulting fitted values for $l_\varphi$ and $l_{SO}$ are discussed in Fig.~1e of the main text.

\end{document}